# EnergyNet in Practice: Long Grid, Short Grids and Mobility-Based Energy Peering

*Parallel deployment, critical-service continuity and the "don't pay twice" principle*

Jonas Birgersson, Max Collins, Marc A. Weiss, Jimmy Chen, Daniel Kammen, Tomas Kåberger, Mark Z. Jacobson, Newsha K. Ajami, Franklin Carrero-Martínez, Michael Menser, Henrik Ny, Lou Riordan, Jill G. Ferguson.

## Abstract

Electrification is planned as a network problem: more demand, more upstream capacity, local resources waiting for reinforcement. We argue that a material part of that bill, in money and time, follows from architecture, not electrification itself, and propose the Short Grid as the deployment unit: a locally controlled building, block or campus whose demand, generation, storage and local market operate behind Energy Routers enforcing an approved connection envelope to the Long Grid. Because the envelope governs exchange, not installed capacity, local resources can grow and many sites can build in parallel at the pace of PV and EV installation, not grid reinforcement. PV and EV batteries dominate today, but the architecture is general: any generation or storage, fixed or mobile, fits behind the same boundary and interfaces. EnergyNet Operators coordinate exchange among Short Grids and with the Long Grid; protection stays local; Vehicle-to-EnergyNet (V2EN) and Mobility-Based Energy Peering (MEP) make mobility-financed batteries accessible within and between them; priority-based operation protects critical services in shortage.

We test it in models, not yet the field: a ten-building neighbourhood in Lund, Sweden; a bus-linked microgrid pair at Stanford and Half Moon Bay; and a proposed 20-site network in Rockaway, New York. Households and services are the largest demand category in Sweden, Denmark, Norway, California and New York (484 TWh/yr); halving their grid imports would be about 242 TWh. Europe's network investment to 2040 is estimated above €1.2 trillion; modelling shows it depends on system design. Negagrid value, following the negawatt, counts network investment avoided, downsized or deferred, net of local substitute costs. "Don't pay twice": build coordinated local capability first; reinforce the Long Grid for what remains.

## 1. From network architecture to repeatable implementation

Electrification is usually planned as a network problem: more demand requires more upstream capacity, and local resources wait for reinforcement before they can connect. Building on EnergyNet Explained [1], this paper asks a different question: which network investments remain necessary once local generation, storage, priority-based operation and mobile support are designed to work together? EnergyNet's central investment proposition is that local electrification need not automatically require a second investment in proportionate upstream capacity. A building, block or campus can instead invest in a Short Grid that coordinates local generation, storage and demand while making better use of its existing Long Grid connection. Compatible vehicle batteries, primarily purchased for mobility, can provide additional flexibility and reserve services. "Don't pay twice" expresses the decision: compare additional network and dedicated storage investment with the full cost of the local capability that could make it unnecessary.

EnergyNet Explained [1] describes the underlying architecture of Energy Routers, local and wide-area energy networks, the Energy Protocol (EP) and operator-scale management. This paper develops its practical implementation: independently deployable Short Grids that cooperate to support additional electrification, maintain critical services and reduce the infrastructure required to do so.

We use Short Grid for the locally controlled deployment domain, from a building to a block or campus. Long Grid denotes the retained national or multinational transmission and distribution system. A Short Grid can retain Long Grid access while managing its own resources. The Long Grid is also distinct from EnergyNet's wide-area network: the former is existing infrastructure; the latter describes coordination and interconnection between EnergyNet domains [1].

The distinction matters causally. A study that gives every alternative the same installed PV, storage and charging portfolio can measure dispatch improvements but overlook the resources that network constraints would have prevented from being commissioned. Short Grids are therefore evaluated both as an operating architecture and as enabling infrastructure. The first question is what can be installed and put into useful service within an existing connection envelope. The second is what the resulting system delivers, and at what total cost.

Ember's The Age of Power identifies coordination and connection as growing constraints on electrification: variable supply, storage and flexible demand can remain poorly matched even as the underlying technologies become affordable [16, pp. 35–36]. Its deployment framing also contrasts modular equipment installation with slower upstream connection processes [16, p. 51]. This motivates the Short Grid question: not only how to operate resources more efficiently, but how to make otherwise constrained resources deployable.

### 1.1 Relationship to adjacent work

The SAFEr Grid Perspective proposes independently controlled, asynchronously interconnected subgrids and energy routing supported by storage [2]. It supplies a complementary theoretical direction: localise control and disturbance exposure without giving up exchange. It also identifies unresolved protection, stability and cyber–physical validation needs. The present paper translates that architectural direction into a deployment and evaluation model linking a physical local domain, an operator role, vehicle interfaces and regional adoption.

The companion paper by Collins et al. [3], referred to here as Quantifying EnergyNet performance, provides the computational framework and numerical examples used in Sections 5 and 6. Its optimisation establishes feasible energy allocation under modelled constraints. The January 2026 Volts conversation [4] is used as a practitioner account of parallel deployment, buffering and the Lund project rationale.

### 1.2 Three contributions

The first contribution is a repeatable deployment model: autonomous local resources behind an enforceable boundary, with interoperable coordination rather than bespoke redesign of the wider network. The second is a V2EN/MEP service model that makes mobility-financed batteries accessible within and between those domains. The third is an assessment of scale: electricity-use examples from Sweden, Denmark, Norway, California and New York show why the built environment is a consequential deployment base, while the Lund, Stanford–Half Moon Bay and SAER/Rockaway[1] cases examine practical operation.

### 1.3 The argument in brief

The paper's argument can be stated in six steps.

1. Buildings are where most electricity is used and where most demand growth will land: more than half of global consumption today and 49% of additional demand to 2030 (Section 8.1).
2. Serving that growth conventionally implies network investment in the trillion-euro range, and published modelling shows that the size of the bill depends on how resources are organised (Section 8.5).
3. A Short Grid lets a building, block or campus add generation, storage and charging behind a controlled connection, so local growth does not require proportionate upstream capacity, and many sites can build in parallel at the pace of local installation (Sections 2 and 2.5).
4. Batteries bought for mobility can supply part of the storage and reserve, within and between Short Grids (Section 3).
5. Three modelled cases test whether this combination delivers service, relief and continuity (Sections 5–7); five reference systems show the scale at stake (Section 8); negagrid value counts the network investment it avoids (Section 8.5).
6. Don't pay twice: build coordinated local capability first and reinforce the Long Grid for what remains (Section 9).

### 1.4 Terms used in this paper

Terms are used as follows; Sections 2 and 3 give the full definitions.

**Short Grid**, a locally controlled electrical domain, from a building to a block or campus, combining demand, generation, storage, a local energy market and a controlled connection to the Long Grid.

**Long Grid**, the retained national or multinational transmission and distribution system.

**Energy Router**, the power-conversion and control device at the Short Grid boundary; it enforces electrical and operational limits and provides ports for local resources and vehicles.

**Energy Link**, an electrical interconnection between Energy Routers, rated separately from the Long Grid interfaces.

**Connection envelope**, the approved import and export power, ramp-rate, power-quality and protection requirements at the Long Grid interface; it governs external exchange, not installed local capacity.

**EnergyNet Operator**, the role that coordinates resources within Short Grids, exchanges between them, and imports, exports and flexibility with Long Grid operators, while protection and essential control remain local.

**Energy Protocol (EP)**, the open protocol through which resources and routers state operating terms, capabilities and requirements using a common resource abstraction; local controllers enforce the electrical limits.

**Vehicle-to-EnergyNet (V2EN)**, integration of compatible vehicles as Short Grid energy resources through bidirectional Energy Router ports and vehicle interfaces, with mobility as their primary service.

**Mobility-Based Energy Peering (MEP)**, transfer of stored energy between geographically separated Short Grids through vehicle journeys; a delayed, capacity-limited path, not a continuous line.

**Critical and flexible demand**, demand maintained whenever physically feasible, and demand that may be reduced or deferred during constrained operation; the basis of priority-based operation.

**Critical-service continuity**, maintaining specified critical services under defined disturbances, with flexible demand reduced or deferred, even when full service is temporarily unavailable.

**Negagrid value**, network investment avoided, downsized or deferred while specified services are maintained, net of the local costs that make the substitution possible.

## 2. The Short Grid as a deployable electrical domain

A Short Grid comprises a defined set of loads, generation, storage, controllable connections and participating owners. Local power conversion and distribution are physical infrastructure, not solely a virtual aggregation of separately connected assets. The Energy Router boundary makes external exchange an explicit operating variable. A larger local PV array need not imply exporting its full nameplate output; a larger charger fleet need not imply simultaneous import at every charger's rating. Local use, storage, sharing, scheduling and, when necessary, curtailment determine the remaining requirement at the Long Grid interface.

Rooftop and parking-area PV and vehicle batteries are the resources that make the case today, because they are inexpensive and are being installed regardless, but the architecture is technology-agnostic. Any generation or storage resource, stationary or mobile, is represented through the same EP resource abstraction and operates behind the same Energy Router boundary and connection envelope: stationary batteries, thermal storage, local combined heat and power, or other generation as it becomes available. The Short Grid's function is the same in each case: to make local resources deployable and coordinated within an existing connection.

[1] SAER, the Rockaway project of [24], is unrelated to the SAFEr Grid Perspective of [2]; the similarity of acronyms is coincidental.

## 2.1 The envelope governs exchange, not installed capacity

The approved external operating envelope specifies maximum import and export power and, where applicable, ramp-rate limits. It also records voltage and power-quality requirements, protection settings and locally enforced protective responses. The aggregate rating of internal resources is not itself the external exchange. Nevertheless, local cable and converter ratings, energy availability and safe operating margins still apply. Where interfaces share a feeder, its aggregate constraint must also be respected; satisfying each port rating alone is insufficient. The connection envelope governs external exchange rather than imposing a one-for-one ceiling on locally installed resources.

## 2.2 Autonomy, operators and the local market

EnergyNet Operators coordinate resources within Short Grids, orchestrate interconnections and exchanges between them, and coordinate imports, exports and flexibility with Long Grid operators (Figure 1). Local resource owners retain authority over asset policies and minimum service requirements. Essential protection and local energy control remain enforceable even if an operator platform or wide-area communications fails. In EnergyNet, EP communicates operating terms and resource states; local controllers and protection enforce electrical limits, owner reserves and critical-service priorities, including when external orchestration is unavailable or issues an incompatible instruction.

The local market enables exchange within local ownership, permission, priority and compensation constraints: energy can be shared within common ownership, exchanged under bilateral terms, or offered through a market. Settlement records source, recipient, quantity and payment. Open interfaces and standards allow multiple vendors and operators to participate without requiring a common equipment supplier [1,5].

## 2.3 Priorities preserve essential service when full service is impossible

An EnergyNet Short Grid supports different service priorities over shared electrical infrastructure. Connected demand is not treated as a single indivisible load: it is classified according to the functions that must be maintained and the activities that can tolerate adjustment. Following the companion simulation framework, this paper distinguishes critical demand, which is maintained whenever physically feasible, from flexible demand, which may be reduced or deferred during constrained operation [3].

When sufficient resources are available, both classes are served. When generation, stored energy or transfer capacity becomes insufficient to meet all requested demand, the operating policy prioritises critical service and adjusts flexible demand. The intended outcome is graceful degradation: reduced overall consumption without an equivalent loss of essential functions. This is selective control of loads and their operating schedules, not an indiscriminate reduction in supply voltage or the same percentage reduction applied to every appliance.

Service priorities must be mapped to independently controllable ports, circuits or compatible devices, with explicit minimum-service requirements and permitted interruption or deferral. The priority policy is therefore implemented locally, not inferred from an aggregate building connection.

Priority-based operation also changes the use of reserves and interconnection. Stored energy can be preserved for the critical-service requirement over the anticipated disturbance, rather than exhausted attempting to maintain unrestricted demand. V2EN and Mobility-Based Energy Peering can extend that service period by providing additional accessible energy.

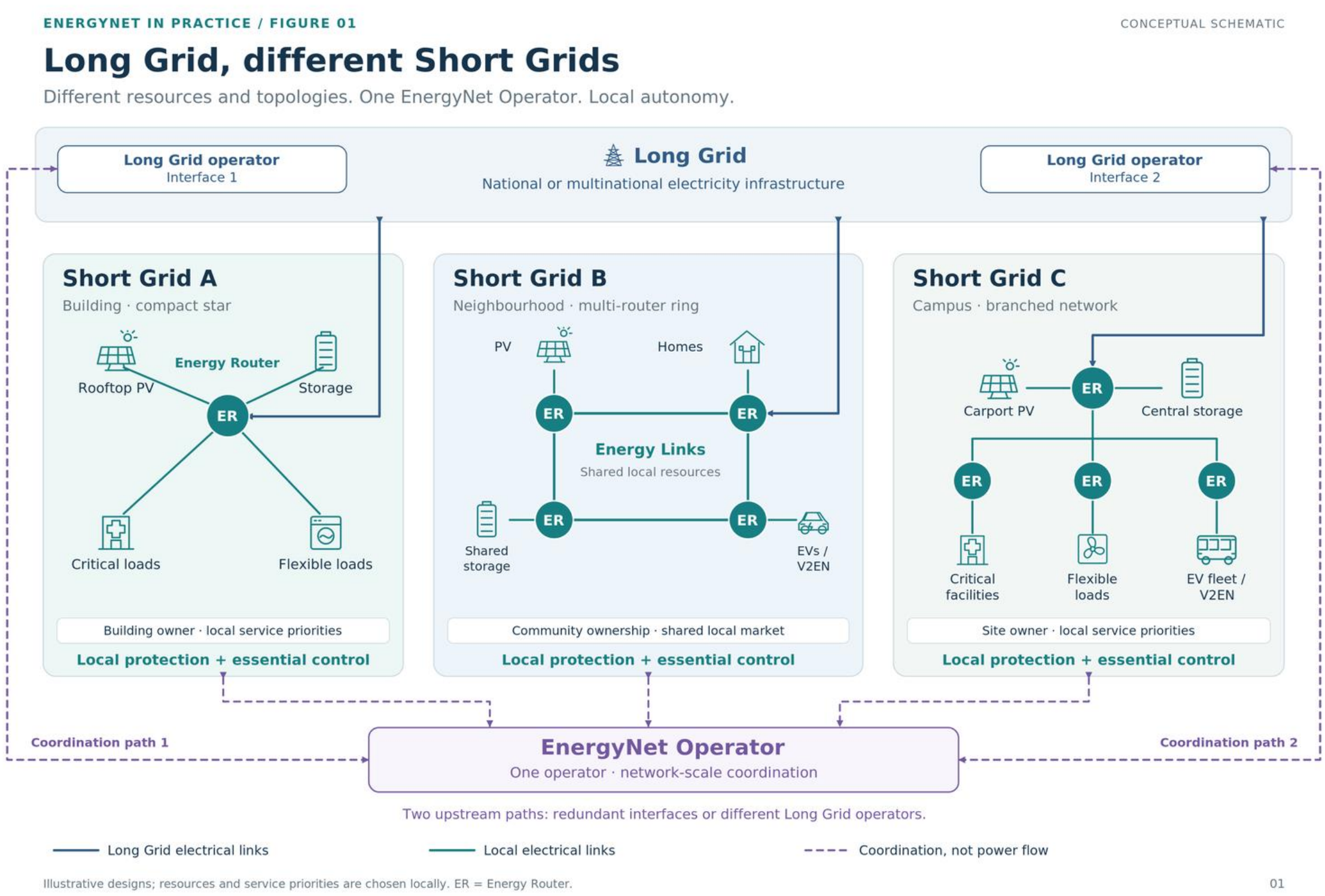

**Figure 1. Long Grid, different Short Grids. Different local resources and topologies share common interfaces and operator coordination. Solid lines are electrical connections; dashed lines are coordination, not power flow. The two upstream coordination interfaces may serve redundancy or different Long Grid operators; they do not establish independent electrical supplies. Local protection and essential control remain within each domain.**

### 2.4 The office LAN precedent: build locally, interconnect at the boundary

An office could install an Ethernet local-area network (LAN) to connect computers, printers and local servers without procuring a telephone-company circuit for each internal exchange. Adding a workstation was an internal network decision, not inherently a request for the carrier to approve another endpoint or redesign the wide-area network. Multiple offices could therefore build in parallel while external service provisioning proceeded separately. Ethernet supplied local packet communication; private IP networking later formalised autonomous internal addressing without per-host coordination with an Internet registry [17,18].

When external connectivity was needed, a router or gateway connected the LAN to a wide-area network and the Internet. The site obtained the necessary upstream service and respected its interface and capacity constraints, but local traffic did not have to traverse that service. A larger internal network did not require a proportional increase in external capacity. The Internet could interconnect independently developed networks instead of becoming a prerequisite for each local network's construction [18,19].

Loss of the external connection did not, by itself, stop local packet delivery: users could still reach a local server or printer if local equipment, power, addressing and required services remained available. Remote applications were unavailable, and a cloud-dependent service was not made local merely by attaching its device to a LAN. RFC 1122 explicitly requires host operation on an isolated network without a gateway [19, §3.3.1.1]; the Volts discussion uses the same local-printer example [4, p. 10].

The Short Grid applies this separation to deployment and operation. Buildings can add local generation, storage and charging behind a controlled boundary while the Long Grid supplies wider exchange. During upstream disruption, local protection, generation and reserves can maintain specified services, with MEP providing later replenishment.

### 2.5 Deployment at the pace of local installation

The time argument follows from the same boundary. Within an approved envelope, the schedule for adding local generation, storage and charging follows local equipment installation rather than the completion of upstream reinforcement and because each Short Grid is commissioned against its own envelope, many buildings and neighbourhoods can proceed in parallel. The Long Grid connection becomes a constraint to respect, not a prerequisite to expand first. Demand growth is arriving on the same timescale. The IEA projects that residential and commercial buildings will account for 49% of additional global electricity demand between 2025 and 2030 [27]. EV charging, counted separately as transport demand, also arrives largely at buildings: 70% of charging energy in the Swedish measurements was supplied at home [7]. The equipment that creates this demand, and the PV and batteries that can serve it locally, is modular and installed on equipment timescales, while upstream connection processes are slower [16, p. 51]. PV adds no load, but under the conventional approach its export creates a connection requirement of its own; the Short Grid envelope governs both directions. Delivering this deployment model requires repeatable engineering, commissioning and approval processes (Figure 2). The investment implication is that local resource growth need not require an equivalent amount of additional upstream connection capacity.

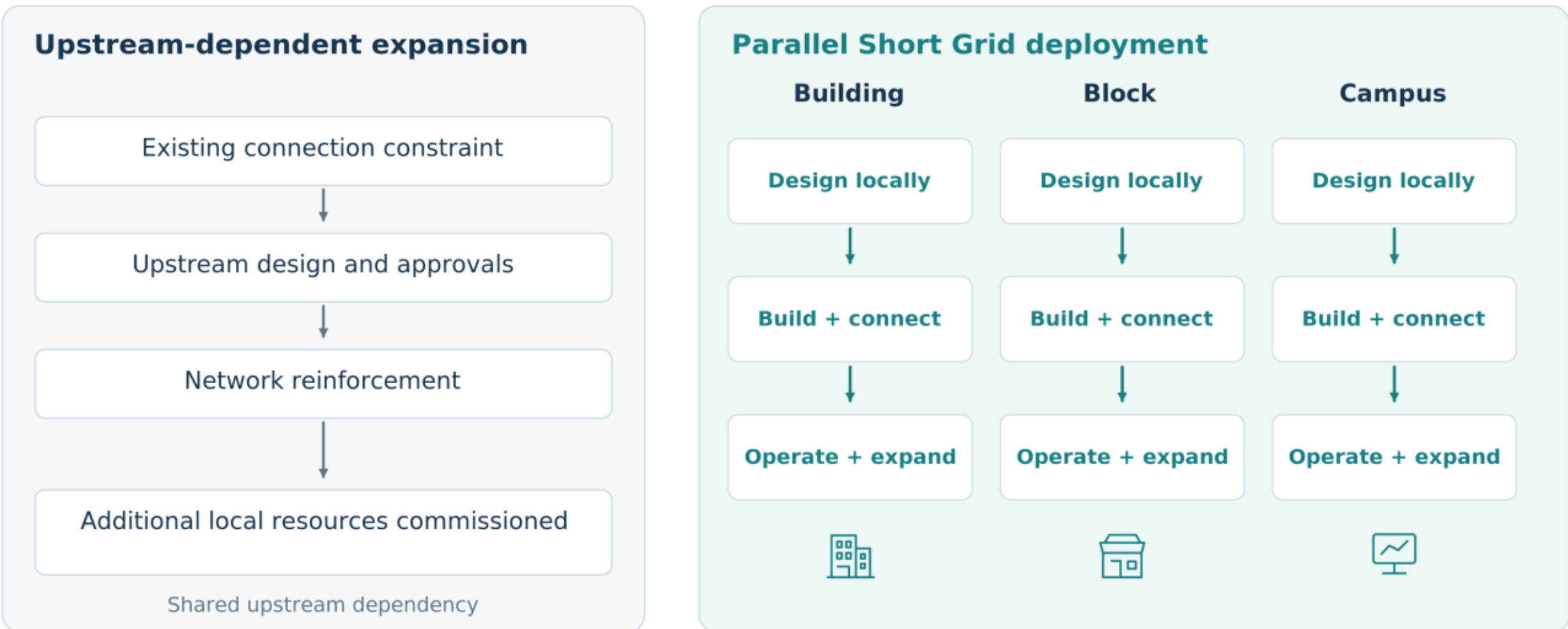

**Figure 2. The two paths differ in adoption pattern: serial, upstream-first expansion versus parallel local deployment. The left-hand path represents projects whose commissioning depends on upstream reinforcement; the right-hand path separates repeatable local work from that dependency. Local engineering and approved connection envelopes still apply. Arrow lengths are not elapsed-time estimates.**

## 3. V2EN and Mobility-Based Energy Peering

V2EN makes a compatible vehicle an energy resource of the Short Grid while retaining mobility as its primary service. Where a suitably rated Energy Router port already provides bidirectional conversion and control, those functions can be reused rather than duplicated in a standalone system. The vehicle interface must still provide compatible voltage and current ranges, protection, isolation as required, session control, metering and authorisation.

### 3.1 One service abstraction, different physical devices

EnergyNet provides the common coordination concept: some functions are native to EP, while others use complementary open standards [5]. The proposed V2EN service profile lets a neighbouring router and a vehicle express energy capabilities and requirements through the same resource abstraction. In EP 1.0, each side restates its operating terms; transfer begins when the terms overlap and the receiving side permits positive intake. Table 1 combines this execution model with the service-level responsibilities needed for V2EN; it is not a claim that every responsibility is a native EP field.

**Table 1. Information and operating behaviour supporting V2EN energy exchange.**

| Service element | Required meaning |
|---|---|
| Identity and authority | Resource, owner, local domain, authenticated peer and permitted actions. TLS 1.3 protects the communication channel; ownership and permissions are handled by the application. |
| Capabilities and operating conditions | Current supply and demand terms, power limits and delivery windows. A transfer identifier distinguishes each exchange episode. |
| Resource constraints | Energy state, mobility and departure requirements, protected service reserves and availability assumptions, applied through locally determined limits. |
| Execution and continuity | Each side states and updates its conditions. Transfer starts when those conditions are compatible and the receiving side permits intake above zero. If communication is lost, the last forecast step remains in effect and each device continues its own operating profile; local electrical limits still apply. |

Metering uses status information and each side's energy ledger for individual time intervals. Financial settlement is handled by the backend. A zero transfer limit or an absent relevant section indicates non-participation; EP does not send an acceptance or rejection message [5].

For example, a receiving resource may express a requirement for 10 kWh between 18:00 and 20:00, with an intake limit of 5 kW. A vehicle may set its supply limit to zero to preserve its next journey; a neighbouring Short Grid may do the same to protect its critical reserve. Transfer can start when the stated conditions overlap and the receiving limit is positive. These values illustrate the common resource abstraction, not a measured transaction.

### 3.2 Storage that is accessible, not merely installed

EV batteries are primarily purchased for mobility, making additional energy services an incremental-use proposition. Their system value depends on connection, location, power, owner consent and energy above protected reserves, not gross fleet capacity [6]. Swedish measurements of 334 EVs found 70% of charging energy supplied at home and a most-frequent parking duration of 14 hours [7]. Parking does not establish connection or permission to discharge. Plug-in assumptions can change estimated availability by more than a factor of two [8].

### 3.3 MEP adds a time-dependent support path

MEP carries stored energy between Short Grids. It is a delayed, capacity-limited path, not a continuous electrical line. We distinguish opportunistic transfer on an unchanged journey from a scheduled support visit that preserves the vehicle's ordinary duty but adds travel. Both can replenish local reserves.

## 4. Evaluation: deployment, operation and maintained service

The evaluation builds on the receding-horizon framework in Quantifying EnergyNet [3], with proposed extensions for vehicle location, energy provenance and time-dependent connection. Its optimisation is a feasibility benchmark, not a reproduction of distributed EP execution or converter dynamics. Lund and Stanford–Half Moon Bay results are attributed to the companion manuscript [3]; the SAER case uses the separately supplied three-figure scenario summary [24]. Neither source's calculations have been independently rerun here.

Two comparisons separate operating value from deployment value. The first holds demand, weather, mobility requirements and installed resources constant and compares dispatch. The second holds the existing connection, commissioning date and required services constant and allows the deployable resource portfolio to differ. The Lund field programme will report against the same definitions:

– annual Long Grid imports against a documented pre-commissioning baseline, over a complete year with consistent initial and terminal storage state;
– coincident peak import at the connection point, aggregated before network peaks are calculated;
– the critical-load fraction sustained and the critical energy delivered during any controlled partition test, with flexible demand deferred or curtailed and reserve depletion recorded;
– V2EN availability as connected and permitted battery energy above protected reserves, not gross fleet capacity;
– conversion losses and travel consumption carried in the energy ledger, with mobile transfers counted once across sites.

Table 2 sets out the evidence status of each result used in this paper once, so that Sections 5–8 can report them plainly. None is field data.

**Table 2. Evidence status of the results used in this paper.**

| Result | Source | Status |
|---|---|---|
| Lund: all critical and flexible demand served in the 72-hour summer and winter cases; summer imports zero | [3, §7] | Simulated, 72-hour windows, field measurement planned |
| Lund: one interface or one ring link lost for 18 h, all demand served; two links lost, 65% of island critical load sustained at 10% reserve, winter median about 92% at 30% | [3, §7] | Simulated, 18-hour partition; 1,000-trial Monte Carlo |

| Result | Source | Status |
|---|---|---|
| Lund: about 60% lower annual imports and 50% lower peak import | Project team | Estimate; field measurement planned |
| Stanford Farm–Half Moon Bay: about 21 kWh delivered over three bus visits; receiving reserve replenished; bus minimum 157 kWh above the 70 kWh mobility reserve | [3, §6] | Simulated, 72 h at 30-minute steps |
| SAER Rockaway: stationary supply 49.1 MWh of a 129.4 MWh critical requirement (38%); 86.4 MWh emergency target met with 700 EVs and 30 buses | [24] | Scenario summary; aggregate energy only, no time-resolved trajectories |
| Five-system 50% import-reduction illustration (Table 5) | [11–14] | Conditional arithmetic, not a forecast |

## 5. Lund: deployable capacity, grid relief and fault tolerance

The Lund configuration described here is under construction at the time of writing (September 2026); measured results will be reported separately. Lund tests the deployment proposition itself: whether a neighbourhood can add PV and storage within an existing connection, and whether shortfalls stay local when links fail. Lund connects the deployment proposition to a commercial neighbourhood design. The Volts account distinguishes an initial two-building technical pilot from a subsequent ten-building programme and describes rooftop PV that the conventional connection would otherwise restrict [4, pp. 34–40].

### 5.1 Configuration and results already available

The Lund deployment comprises ten buildings and 278 homes, using a ten-router closed ring represented in Quantifying EnergyNet [3, §7]. The model's annual demand is 650 MWh, with 500 kWp PV and 1 MWh distributed battery storage; each battery has a 50 kW rating. The ten Energy Links are rated at 100 kW, and two Long Grid interfaces, at R1 and R6, are each rated at 300 kW import and 150 kW export.

**Table 3. Lund results reported in Quantifying EnergyNet. These are simulated energy-service results, not field availability.**

| Case in [3] | Reported outcome and boundary |
|---|---|
| 72-hour summer / winter operation | All critical and flexible demand served. Summer imports are zero; winter remains grid-supported. |
| One grid interface or one ring link lost for 18 h | All demand served through the surviving supply and transfer paths. The router itself remains intact in the grid-interface case. |
| Two links lost; R4 & R5 isolated for 18 h | Other eight routers fully served. With 10% protected reserve, the text reports 65% of island critical load sustained. |
| Reserve sweep and conditional Monte Carlo | About 30% reserve enables full critical service in the deterministic case. At that reserve: full summer service in 1,000 trials; winter median sustained critical fraction about 92%. |

The modelled energy-service result is alternative supply and localisation of shortfalls: a lost link need not disconnect the whole domain, and a partition can reduce service locally while the remainder stays supplied. Physical disturbance containment still requires boundary and protection validation. The limit is also explicit. The reported winter median is a fraction of critical load maintained during the imposed partition, not 92% annual reliability.

### 5.2 Annual imports and peak load: project estimates

The basis of these estimates is as follows. Annual generation of the specified 500 kWp array is approximately 480–500 MWh using the PVGIS site data adopted in [3, §7.1] some 73–77% of the 650 MWh annual demand and the ~60% import-reduction figure is the project team's extrapolation of the seasonal dispatch behaviour of [3] (local use first, distributed storage, inter-building transfer over the ring) across the full demand profile. The ~50% peak-reduction figure is consistent with comparable Swedish district developments: feasibility studies for the Tamarinden district in Örebro, a new-build neighbourhood combining local generation, storage and inter-building energy sharing, project a 50% reduction in power demand (effektbehov) [40]. These projections are distinct from the representative 72-hour simulations above; the full-year comparison, baseline definition and supporting peak series will be reported against the metrics of Section 4.

### 5.3 From pilot to programme: deployment sequence and regulatory basis

The deployment model implies a repeatable sequence, which the Lund programme is designed to follow. The connection envelope, in practice the connection agreement at the existing Long Grid interface, is settled with the distribution network operator; Energy Routers, PV and storage are installed and commissioned against that envelope; an EnergyNet Operator is onboarded; and further buildings are added within the envelope in parallel rather than in series. The Volts account describes the first stage as a two-building technical pilot preceding the ten-building programme [4, pp. 34–40]. The evidence sequence is equally deliberate: this paper establishes the impact potential in models; the field measurements will be compared with the simulated results in Table 3 against the metrics of Section 4; scaling follows verification.

The regulatory precondition for building behind the boundary already exists in Sweden. Under the exemption regulation for non-concession networks, the IKN regulation (2007:215), the internal networks it lists may be built and used without a network concession [32,33]. The amendments in force from 1 January 2022 widened those exemptions to include sharing of electricity between a property owner's buildings within stated limits, and further changes took effect on 14 January 2025 [34,35]. A property owner therefore does not need the grid company's consent for its internal network; the connection to the concession grid remains governed by the connection agreement, which is the envelope of Section 2.1, and the Energy Markets Inspectorate issues binding decisions on whether a planned network is exempt [33]. EU law provides the corresponding rights: active customers and citizen energy communities under Directive (EU) 2019/944, renewables self-consumers acting jointly under Directive (EU) 2018/2001, and the right to energy sharing inserted as Article 15a of Directive (EU) 2019/944 by Directive (EU) 2024/1711, with transposition of the energy-sharing provisions due by 17 July 2026 [36–38]. Transposition into specific national rules remains uneven, and the Commission's 2026 recommendation on energy sharing addresses consistent implementation of the right [39]. Where those rules exist, the parallel-deployment pathway of Section 2.5 is legally open; where they do not, transposition is the binding constraint, not technology.

## 6. Stanford Farm–Half Moon Bay: a mobile support path

The Stanford Farm–Half Moon Bay case tests anticipatory mobile support: whether a vehicle on its ordinary duty can replenish a remote site before its reserve is exhausted. The paired-microgrid case in Quantifying EnergyNet [3, §6] provides a concrete application of mobile energy support. Its Stanford Farm–Half Moon Bay configuration links two separately located energy systems through a campus electric bus. No direct electrical line connects them in the model. Operationally, an anticipated deficit is expressed as a time-windowed requirement, the vehicle is checked against its duty and reserve, and local reserves must carry the receiving site until the vehicle arrives. Stanford's public material documents farm electrification and a separate research programme coordinating solar, storage and campus bus operations [9,10]. The combined dispatch below is the Quantifying EnergyNet modelled application of that infrastructure, with site assumptions stated explicitly.

**Table 4. Parameters adopted in Quantifying EnergyNet for the Stanford Farm–Half Moon Bay example.**

| Parameter | Farm / depot | Half Moon Bay | Mobile resource |
|---|---|---|---|
| PV capacity | 35 kW | 20 kW | – |
| Battery capacity | 45 kWh | 60 kWh | 350 kWh nominal |
| Power / energy limits | Battery: 15 kW | Battery: 20 kW | 70–315 kWh operating range |
| Grid / charging interface | Grid assumed; 2 × 11.5 kW | Off-grid; 50 kW receiving | 30 kW depot charging |

The receiving site experiences PV output at 20% of the clear-day profile. Forecasts identify declining storage and trigger a preventive visit after the bus completes its normal campus duty. It then returns to the depot and recharges overnight. Quantifying EnergyNet reports about 21 kWh delivered over three visits and minimum bus energy of about 157 kWh, above the 70 kWh mobility reserve. Without support, the receiving battery approaches depletion by the end of the simulation; with support, its reserve is replenished [3, §6.2].

### 6.1 What the reported case establishes

The bus retains its transport duty, but the described support visit follows that duty. Furthermore, Half Moon Bay is already off-grid in the model, while depot charging remains available. The case demonstrates anticipatory support during a local generation deficit, not continued service after simultaneous loss of all charging supplies.

## 7. SAER in Rockaway: shared resources and mobile replenishment

SAER tests the share–prioritise–replenish response at district scale during a prolonged regional outage. The SAER proposal extends the analysis from an individual neighbourhood and a pair of microgrids to an interconnected energy system across Rockaway, New York. The supplied design comprises 20 Short Grids, 83 Energy Routers and 28 inter-site connections (Figure 3). Local networks with different configurations are connected so that energy can be shared between sites, rather than treating each site as an isolated generation and backup requirement [24].

### 7.1 Share, prioritise and reinforce

The outage response combines three capabilities (Figure 4). First, the stationary EnergyNet shares available solar generation and stored energy across connected sites. Second, emergency operation protects priority services while reducing demand for other activities. Third, mobile resources replenish the network when stationary energy is insufficient. Fixed links distribute available energy; priority-based operation defines the service requirement to be protected; MEP brings additional energy across the boundary of the affected area [24].

### 7.2 A 48-hour winter outage

The supplied stress-test summary considers a lowest-solar winter period lasting 48 hours, with the regional grid unavailable (Figure 5). The stationary system supplies 49.1 MWh against an original critical requirement of 129.4 MWh, approximately 38% coverage. Emergency operation reduces the target to 86.4 MWh; stationary supply is then reported as 49.0 MWh, or 56.7% of that target [24].

The rise from 38% to 56.7% is principally a change in the protected service-energy requirement, not additional generation. The original target falls by 43.0 MWh, approximately 33.2% using the displayed values.

The mobile configuration adds 700 EVs and 30 buses with one recharge rotation. Vehicles recharge beyond the outage boundary and return. The summary reports 86.4 MWh supplied against the 86.4 MWh emergency-service target, with 22.0 MWh labelled as returned energy. Under the stated scenario, the combined resources meet that reduced energy target.

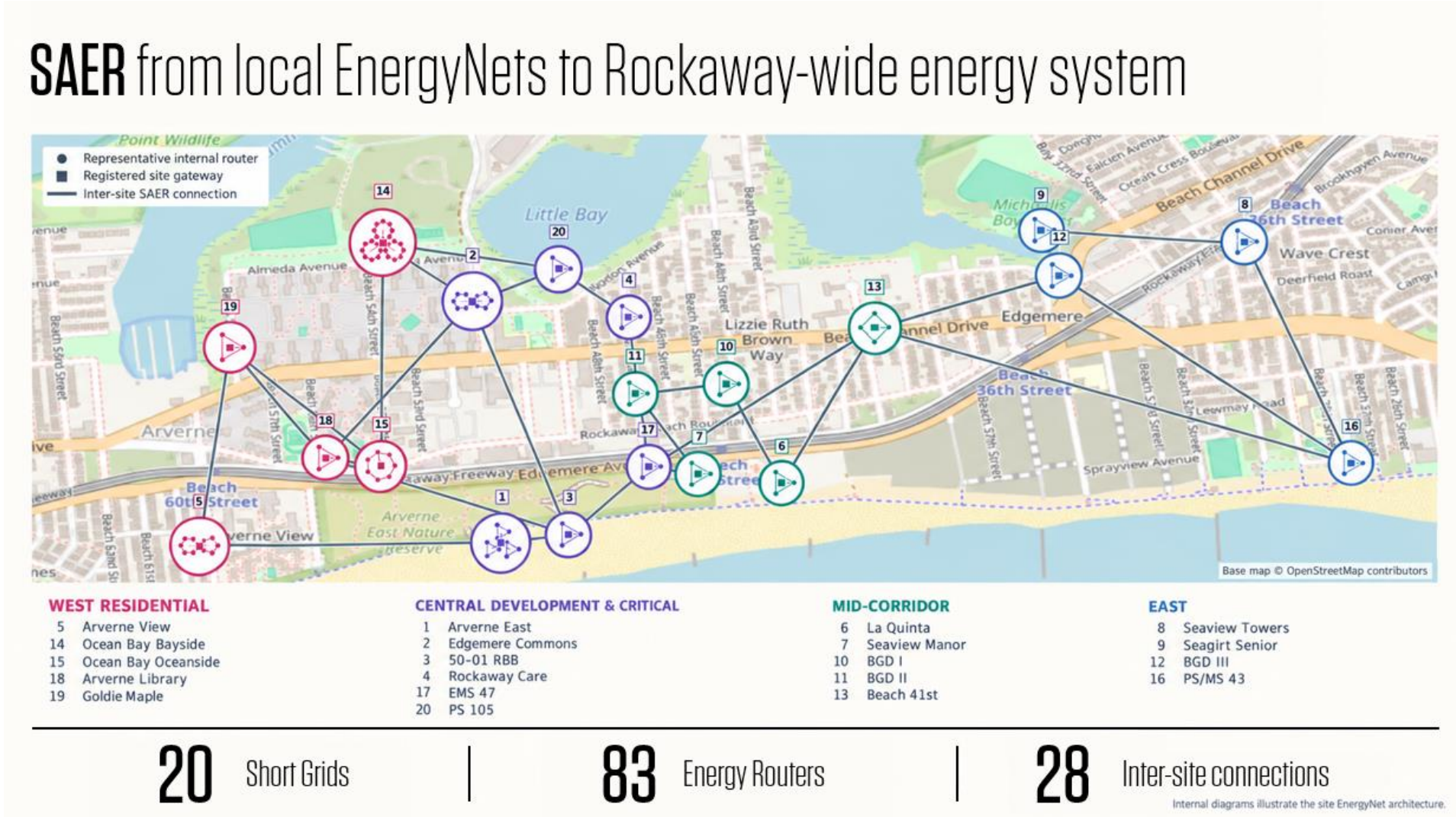

**Figure 3. Proposed SAER network in Rockaway: 20 Short Grids, 83 Energy Routers and 28 inter-site connections. Internal symbols illustrate site configurations; links represent inter-site connections [24]. Base map © OpenStreetMap contributors.**

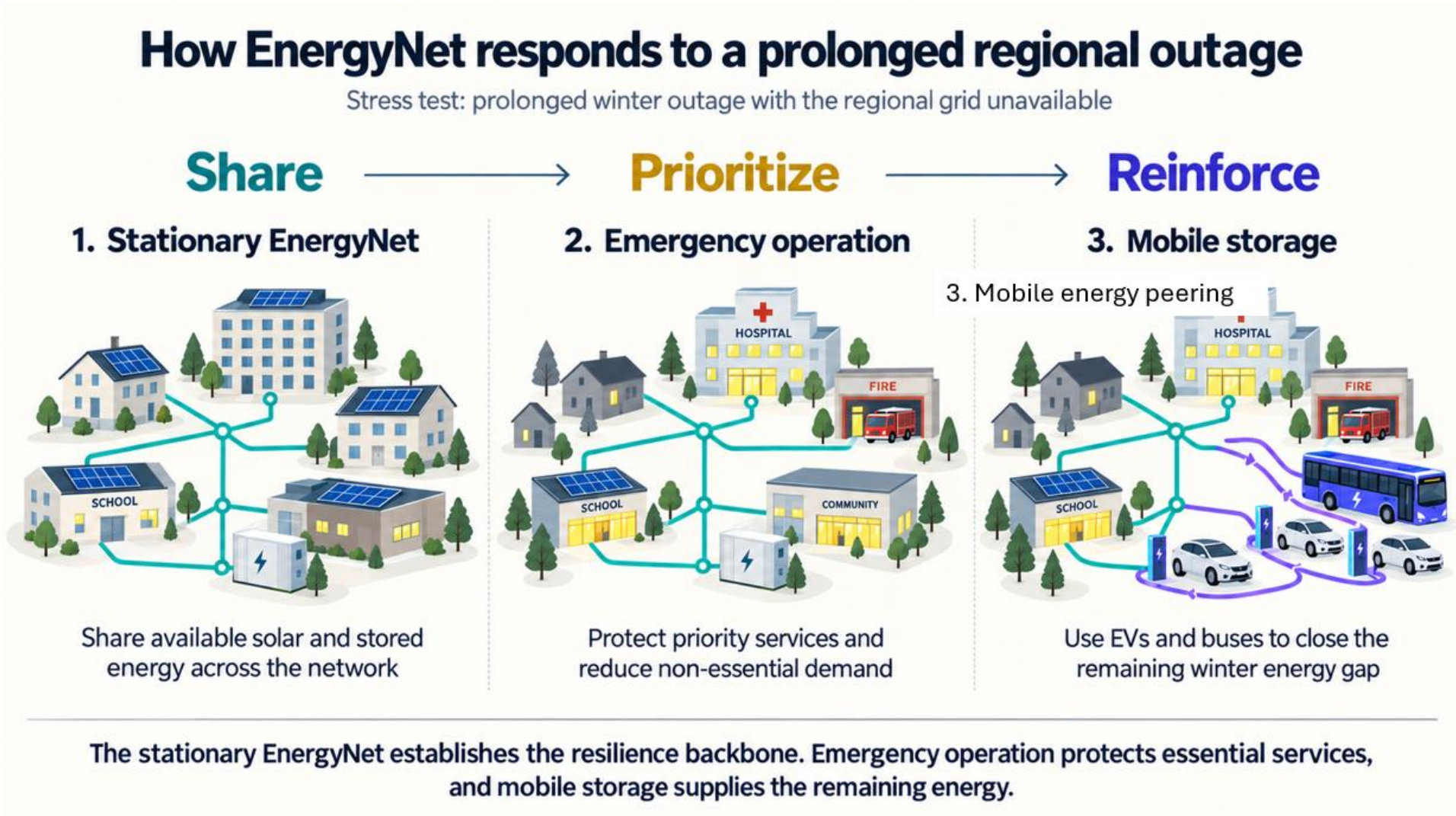

**Figure 4. SAER's proposed response to a prolonged regional outage: share available energy, prioritise essential services and reinforce local reserves with mobile storage. The illustration describes the operating concept, not a fault-protection test [24].**

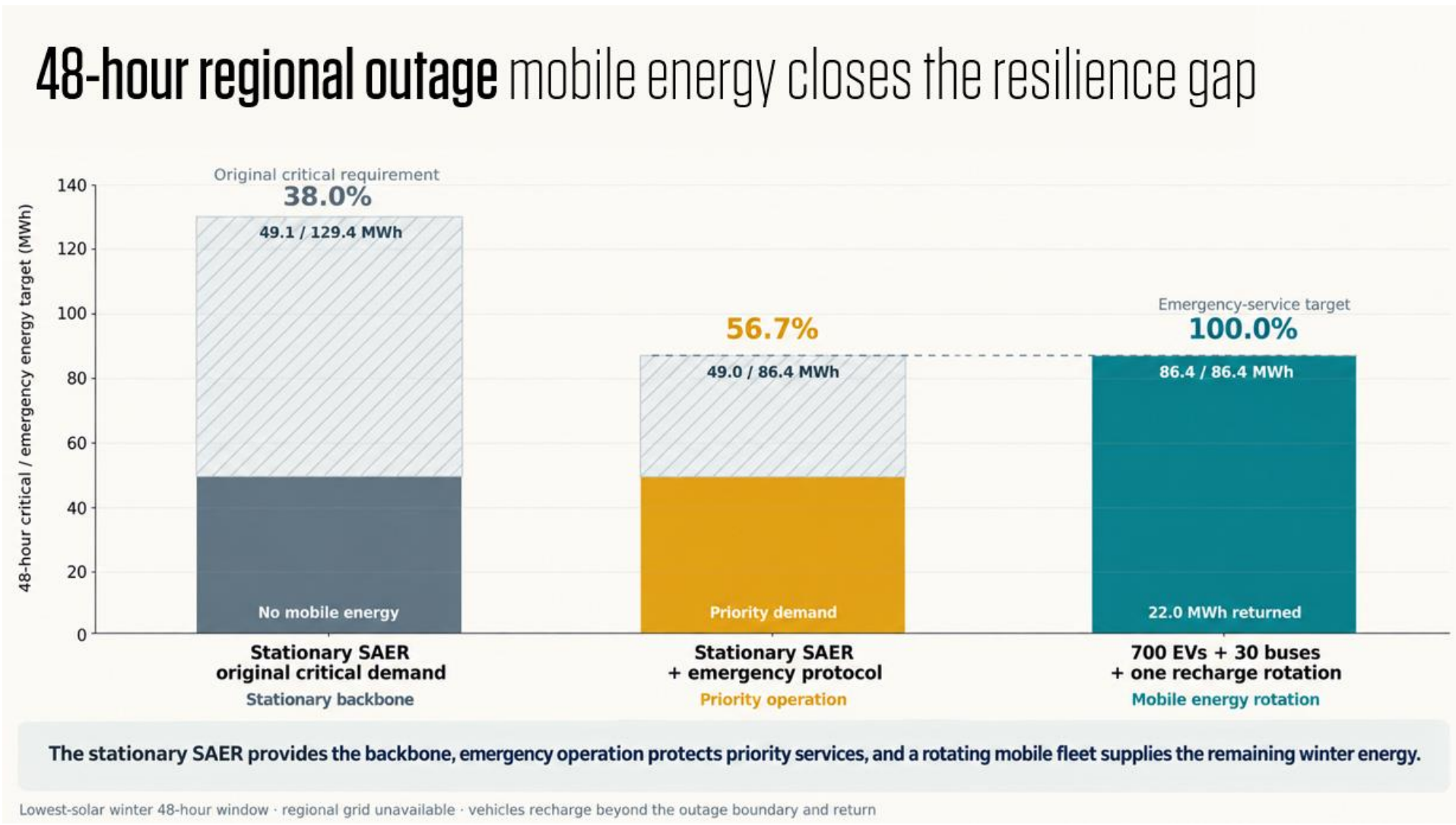

**Figure 5. SAER's 48-hour energy-coverage summary. The first bar uses a 129.4 MWh original critical requirement; the second and third use an 86.4 MWh emergency-service target. The fleet comprises 700 EVs and 30 buses with one recharge rotation outside the outage boundary. Percentages denote energy coverage, not annual availability [24].**

The result illustrates replenishable reserves: the affected network is not limited to the energy held locally when the outage starts. Its coverage depends on vehicles, routes, external charging and receiving interfaces remaining available.

## 8. The built environment: global scale, electrification and "don't pay twice"

The local cases examine how coordinated resources can deliver electricity, mobility and critical services. The wider investment question is whether deploying that capability across the built environment can materially reduce the infrastructure required for electrification. The significance lies in three connected characteristics: the magnitude of building-associated demand, the availability of technologies capable of decarbonising much of its operation, and the opportunity to deploy those technologies without proportionate expansion of upstream networks.

### 8.1 The largest demand category, everywhere it is measured

Two levels of evidence are used here, with different jobs. The first establishes that building-associated demand is the largest electricity-demand category wherever it is measured, not a regional peculiarity. The IEA identifies buildings as accounting for more than half of global electricity consumption [25]. Eurostat reports approximately 1394 TWh of household and commercial/public-service electricity consumption in the EU in 2023, around 60% of the reported sector total [26]. In the United States, residential and commercial customers together accounted for 73.8% of electricity sales in 2024 [14]. These figures frame the opportunity.

The opportunity concerns future demand as well as existing consumption. In Electricity 2026, the IEA projects that residential and commercial buildings will account for 49% of additional global electricity demand between 2025 and 2030 [27]. The proposed Short Grid domain therefore addresses both a large existing demand base and the principal location of further electrification and demand growth, much of which arrives on equipment-installation timescales (Section 2.5).

The second level is the paper's own number. Table 5 applies the household-plus-services proxy to five reference systems: Sweden, Denmark and Norway, and the US states of California and New York, which host the cases in Sections 5–7 and publish comparable sector statistics. The category is the largest in all five, and the range of shares is itself informative: from 51% in Norway, where energy-intensive industry takes a large share of consumption, to 87% in New York, where it does not. The built environment is the largest category whether industry is heavy or light.

**Table 5. Building-associated electricity demand in five reference systems and a conditional 50% import-reduction illustration.**

| Area / year | Households + services (TWh/y) | Share of stated total | 50% × full proxy (TWh/y) |
|---|---|---|---|
| Sweden / 2024 | 75.9 | 60.5% of 125.3 | 37.9 |
| Denmark / 2023 | 19.2 | 63.1% of 30.5 | 9.6 |
| Norway / 2024 | 66.5 | 51.4% of 129.3 | 33.3 |
| California / 2024 | 201.0 | 81.8% of 245.7 | 100.5 |
| New York / 2024 | 121.9 | 86.8% of 140.5 | 61.0 |
| Combined illustration | 484.5 | Not a harmonised regional balance | 242.3 |

Sources and boundaries [11–14]: Sweden combines households and services; its final-use denominator includes separately estimated self-consumed solar not allocated to these sectors. Denmark uses climate-corrected household and service-sector consumption. Norway combines households and tertiary industries. California and New York use residential and commercial retail electricity sales, excluding on-site self-supply. Years and classifications differ; calculations use the underlying unrounded values.

### 8.2 A conditional illustration of grid relief

Applying a 50% reduction to half of this demand proxy corresponds to approximately 121.2 TWh/year; applying it to the entire proxy corresponds to 242.3 TWh/year. These are conditional scale illustrations, not regional forecasts. The illustration is confined to these five systems, whose boundaries are documented above; it is deliberately not applied to the EU or US totals cited earlier, because heating fuels, climate, PV yield and vehicle adoption differ too much for one percentage to be meaningful across them. Achievable reductions depend on eligible sites, local generation, demand profiles, storage availability and adoption. Their significance is nevertheless clear: substantial reductions in the grid dependence of this demand category would change the operation of the wider electricity system, not merely improve isolated buildings.

### 8.3 A decarbonisation opportunity comparable in scale to major industries

Building operations are also a substantial emissions source. UNEP and GlobalABC report operational emissions of 9.9 $GtCO_2$ in 2024, including emissions associated with supplied electricity and heat [28]. On a narrower direct-emissions basis, the IEA's 2024 sector charts place building operations at approximately 2.8 $GtCO_2$, comparable with iron and steel production at approximately 2.7 $GtCO_2$ and cement production at approximately 2.3 $GtCO_2$ [29]. The latter values are rounded readings of the published charts. Building operations therefore constitute a decarbonisation challenge comparable in magnitude to either major industry even before their purchased-energy emissions are included.

The difference lies in the remedy. The IEA's September 2026 Electrification report estimates that approximately half of residential fuel consumption could already be electrified competitively under current technology costs, energy prices and financial conditions [30]: the equipment exists. Grid relief, avoided expenditure and avoided emissions remain related but distinct outcomes. A 50% reduction in Long Grid imports is not a 50% reduction in emissions; carbon performance requires its own assessment of displaced fuels, the emissions intensity and timing of electricity, conversion losses and the embodied impacts of additional equipment.

For EnergyNet, the resulting proposition is a deployment proposition: make otherwise viable electrification possible where connection constraints, coordination requirements and additional infrastructure costs impede it.

### 8.4 Turning local resources into deployable capability

The Short Grid brings together demand, rooftop and parking-area PV, stationary storage and compatible vehicle batteries within a controlled local domain. Its proposed value is not solely improved dispatch after installation. It is also the additional generation, charging and storage that can be commissioned within an existing Long Grid connection envelope. Where local coordination resolves the binding constraint, the deployment schedule can follow local equipment installation rather than the completion of upstream reinforcement.

Two economic mechanisms support this approach. Cross-market component learning can lower the cost of equipment, as discussed in Ember's The Age of Power [16]. Separately, V2EN enables additional use of batteries whose primary investment is justified by mobility. This is also a potential complement to industrial decarbonisation. Low-emissions steel production can require additional electricity, including for hydrogen production [31]. Where Short Grid deployment releases usable capacity at the relevant network locations and times, it can help accommodate such demand. The argument is not buildings instead of industry, but more effective use of shared infrastructure across both.

### 8.5 From negawatts to negagrid value: part of the network bill is a design choice

The financial opportunity extends beyond purchasing fewer kilowatt-hours. The European Commission estimates requirements of €730 billion for distribution and €477 billion for transmission development by 2040 [20].

Published modelling demonstrates that network requirements depend materially on how electrification is organised. Eurelectric's Grids for Speed estimates that combined planning, asset-management and flexibility strategies could reduce annual distribution investment from approximately €67 billion to €55 billion across the EU27 and Norway over 2025–2050. Grid-friendly flexibility contributes approximately €4 billion or more in annual net reductions after the modelled network adjustments, excluding flexibility activation payments [21]. California's Public Advocates Office identifies $5–18 billion in potentially avoided distribution upgrades through 2040 from managed EV charging across the three largest investor-owned utilities, excluding implementation costs [22]. These are differently scoped studies, not additive estimates or EnergyNet-specific savings. They establish that part of the expansion bill depends on system design and operation, not simply on the amount of electrification.

Following Amory Lovins' least-cost service perspective [15], this paper uses *negagrid value* for network investment avoided, downsized or deferred while maintaining specified service requirements, acknowledging the earlier 'nega-grid' terminology of Blarke and Jenkins [23]. Electricity consumption can increase as fossil-fuel services are electrified while the additional network investment required to support them decreases relative to a conventional alternative.

Net *negagrid value* is the present-value difference between network-investment pathways with and without Short Grids, less the incremental local costs required to provide the substitute service.

The practical test is whether local capability removes the constraints that would otherwise trigger a particular upgrade. The comparison must show a constraint removed rather than shifted elsewhere, with equivalent electricity, mobility and resilience services; a lower tariff bill is not by itself a reduction in societal infrastructure cost, and curtailment of required service is not a saving. **"Don't pay twice"** is therefore an investment principle, not a promise of costless infrastructure: use the grid already financed and batteries primarily purchased for mobility more productively and build local capability where it makes additional network or dedicated backup investment unnecessary.

## 9. Conclusion

The built environment is not a peripheral opportunity for grid relief. Buildings use more than half of the world's electricity and are projected to account for 49% of additional demand between 2025 and 2030 [25,27]. In the five reference systems examined here, households and services form the largest demand category, from 51% of electricity use in Norway to 87% in New York, together some 484 TWh a year (Table 5) [11–14]. Halving Long Grid imports across that base would correspond to roughly 242 TWh annually: an illustration of scale, not a forecast, but one that places Short Grids at the centre of the system rather than its edge. In direct emissions, building operations are also a decarbonisation task comparable to steel or cement, with one difference: much of their demand can be met by electrical equipment available today [29,30].

Serving that demand by conventional expansion is expensive: the European Commission's estimates for distribution and transmission development to 2040 exceed €1.2 trillion [20]. That bill is not fixed by the quantity of electricity consumed. Eurelectric's modelling shows that planning and flexibility strategies could reduce European distribution investment from about €67 billion to €55 billion a year, and California modelling identifies $5–18 billion in distribution upgrades that managed EV charging alone could avoid [21,22]. None of these figures is an EnergyNet saving. They show that part of the expansion bill is a consequence of system design, and system design is a choice.

EnergyNet's proposition is that local investment can do part of the work that network expansion would otherwise do, and do it sooner, because Short Grids deploy in parallel within existing connection envelopes. Short Grids provide the deployment domain, EnergyNet Operators coordinate cooperation between domains, V2EN and MEP make mobility-financed batteries accessible within and between them, and priority-based operation directs constrained resources to the services that must remain available. The Lund, Stanford–Half Moon Bay and SAER cases test the combination at neighbourhood, paired-site and district scale; the results are simulated energy-service outcomes and scenario summaries, not field data [3,24]. Negagrid value, following Lovins' negawatt logic [15,23], is the accounting: network investment avoided, downsized or deferred, net of the local costs that make the substitution possible. What matters is the total cost of delivering the specified services, not which side of the connection the investment sits on.

The evidence sequence is deliberate. This paper establishes the impact potential in models: what coordinated Short Grids could save in money and time if the architecture performs as simulated. The Lund field data will test the simulated results against measurement, using the metrics of Section 4. Scaling follows verification, and in Sweden the regulatory basis for building behind the connection already exists (Section 5.3).

The investment decision framed in Section 1 thus becomes a design rule. Once local generation, shared storage, priority-based operation and mobile support are designed to work together, which additional investments remain necessary? Don't pay twice: build the local capability first, and reinforce the Long Grid only for what remains.

## Use of AI-assisted tools

ChatGPT (OpenAI) was used to assist, source checking and language editing. Claude (Anthropic) was used for structural review and for drafting and editing the abstract. AI image-generation tools also assisted the preparation of conceptual artwork. The human authors retain responsibility for technical accuracy, source attribution, interpretation and final approval.

## Data and code availability

Lund and Stanford–Half Moon Bay results are attributed to the companion manuscript [3] and the SAER configuration and energy-coverage summary to [24]; data and code for the modelled cases are held by their authors, and their availability will be stated with the public release of [3]. Regional arithmetic uses the public statistical sources and boundaries stated in Table 5; the California and New York figures reproduce from EIA Table F30 [14]. The Lund deployment is under construction. Pre-commissioning baseline and post-commissioning measurements of Long Grid imports and coincident peak will be reported separately against the metrics defined in Section 4.